\documentclass[12pt]{article}
\usepackage{amssymb,amscd,amsmath,amsthm,eucal}
\usepackage{latexsym,amstext}
\usepackage{epsfig}

\begin{document}

{\Large\bf Approximation methods for the calculation of eigenvalues in ODE with periodic or anti periodic boundary conditions: Application to nanotubes}

\vskip1cm

\centerline{M. Gadella$^\dagger$, L.P. Lara$^\ddagger$, J. Negro$^*$}

\bigskip

{$^\dagger$Departamento de F\'{\i}sica Te\'orica, At\'omica y
\'Optica and IMUVA, Universidad de Va\-lladolid, Paseo Bel\'en 7, 47011 Valladolid, Spain,
manuelgadella1@gmail.com.\\ $^\ddagger$Departamento de F\'isica, FCEIA, UNR, Av. Pellegini 250, 2000 Rosario, Argentina,
lplara2014@gmail.com.\\ $^*$Departamento de F\'{\i}sica Te\'orica, At\'omica y
\'Optica and IMUVA, Universidad de Va\-lladolid, Paseo Bel\'en 7, 47011 Valladolid, Spain, jnegro@fta.uva.es}

\begin{abstract}

We compare three different methods to obtain solutions of Sturm-Liouville problems: a successive approximation method and two other iterative methods. We look for solutions with periodic or anti periodic boundary conditions. With some numerical test over the Mathieu equation, we compare the efficiency of these three methods. As an application, we make a numerical analysis on a model for carbon nanotubes. 

\end{abstract}

\section{Introduction}

The objective of the present article is double. On the first side, we compare three different iterative methods to determine the eigenvalues of a Sturm-Liouville type problem with either periodic or anti-periodic boundary conditions. As a second goal, we apply these algorithmic ideas to a numerical analysis in a situation that appears in the physics of nanotubes. For this analysis, we use a model, which has been already considered by Jakubsk\'y et al. \cite{JKN} and that motivates our research. In particular, the explicit form of the equation to be solved as well as the use of periodic and anti-periodic boundary conditions for its solutions has its justification on this model.  

On its more general form,  the type of systems on which we focus our attention is primarily given by  two ordinary differential equations such as,

\begin{eqnarray}\label{1}
y'(x)=f_\lambda(x,y(x),z(x))\,,\qquad z'(x)=g_\lambda(x,y(x),z(x))\,,
\end{eqnarray}
where the indetermined functions $y(x)$ and $z(x)$ are defined on a given interval $[0,p]$, with $p<\infty$. The tilde denotes derivation with respecto to $x$. The functions $f_\lambda$ and $g_\lambda$ are given data and depend on the variables $x$, $y$, $z$ and linearly on the eigenvalue $\lambda$.  As functions of $x$, they do not have singular points on the open interval $(0,p)$. 

In addition, we impose to solutions, $y(x)$ and $z(x)$, of system (\ref{1}) to satisfy the following boundary conditions:

\begin{equation}\label{2}
y(0)=\alpha y(p)\,,\qquad z(0)=\alpha z(p)\,, \qquad \alpha=\pm 1\,.
\end{equation}

Solutions satisfying (\ref{2}) with $\alpha=1$ or $\alpha=-1$ are called {\it periodic} or {\it anti-periodic}, respectively. 

Thus, one of the objectives of the present work is the discussion of simple and easily applicable methods for the determination of approximate eigenvalues for the Sturm-Liouville problem described so far with particular emphasis to their application to situations that have appeared in physics.   

As an example of these applications, we propose the approximate determination of the energy bands of the nanotube model in \cite{JKN}. This is essential if we want to go beyond the limited situation described in \cite{JKN}, which is exactly solvable. We propose an algorithmic approach for this purpose.  

It is certainly true that there already exist computational methods for problems like this one described here, among which we should mention the finite difference method \cite{B,G,EL}, matrix methods \cite{RA}, the shooting method \cite{IS} and those based on the Runge-Kutta method \cite{RK,RK1}. Nevertheless and motivated by conceptual and operational simplicity, we introduce here three other methods that fit with our aim. They are based in very elementary concepts and are very simple to handle, with a quite easy numerical implementation, which in addition give reasonably good results.   We start with a successive approximation method (SAM) and then give two others which may be classified as iterative methods. As a matter of fact, all three are iterative, although we keep this terminology for the two latter in order to underline the way in which we integrate the differential system. 

This paper is organized as follows: In Sections 2 and 3, we describe the successive approximation and the two iterative methods, respectively. In Section 4, we comment on the convergence of one of our methods. Section 5 is devoted to a numerical comparison between the methods. The application to carbon nanotubes is introduced in Section 6. We include some concluding remarks.  

\section{The successive approximation method (SAM)}

This method will not intend to study equation (\ref{1}) with (\ref{2}) in its full generality, but it will refer to  only one equation with the form:

\begin{equation}\label{3}
y''(x)+(\lambda a(x)-V(x))y(x)=0\,,
\end{equation}
where $a(x)$ and $V(x)$ are know functions. This is a particular case of (\ref{1}), if we choose $z(x):=y'(x)$. Then, both $y(x)$ and $y'(x)$ show identical boundary conditions. 

Note that $\lambda$ is positive whenever $a(x)$ and $V(x)$ be positive. To show this, let  us multiply (\ref{3}) by $y(x)$ and then, integrate by parts the term $\int_0^p y''(x)\,y(x)\,dx$ taking into account the boundary conditions given by (\ref{1}) (with $y'(x)\equiv z(x)$).  This gives:

\begin{equation}\label{4}
\lambda=\frac{\int_0^p [y'^2(x)+V(x)y^2(x)]\,dx}{\int_0^p a(x)y^2(x)\,dx}\,,
\end{equation}
which proves our claim. 

Equation (\ref{4}) suggest the following method of successive approximations: Take the initial values (that may be also looked as boundary conditions) $y(0)=1$ and $y'(0)=0$ and an initial value for $\lambda$, say $\lambda_0$. Assume that we can solve (\ref{3}) under these conditions. We obtain a solution called $y_0(x)$. Then, use this solution in (\ref{4}) to obtain a new value for $\lambda$, that we shall denote by $\lambda_1$. With this value and using the same initial conditions, we solve (\ref{3}) to obtain a solution, $y_1(x)$ and so on. The $k$-th iteration can be written as:

\begin{eqnarray}\label{5}
y''_k(x)+(a(x)\lambda_k-V(x))y_k(x)=0\,,\qquad y_k(0)=1\,, \quad y'_k(0)=0\,,\nonumber\\[2ex]
\lambda_{k+1}=\frac{\int_0^p (y'^2_k(x)+V(x)y^2_k(x))\,dx}{\int_0^p a(x)\,y_k^2(x)\,dx}\,, \qquad k=0,1,2,\dots,n\,.
\end{eqnarray}

The number $n$ of iterations is usually determined by the desired precision on the evaluation of the eigenvalue $\lambda$. For instance, giving a $\delta>0$ such that $|\lambda_{k+1}-\lambda_k|<\delta$ for $k$ sufficiently large. In general terms, we need numerical methods in our calculations. 

We need a procedure to determine the value $\lambda_0$, which is usually named as the {\it seed}. Let us consider the equation 

\begin{equation}\label{6} 
y''(x)+\omega^2 y(x)=0\,,\;\; \omega^2:=\langle a(x)\rangle\lambda_0-\langle V(x)\rangle\,; \;\; \langle f(x)\rangle= \frac1p \int_0^p f(x)\,dx\,.
\end{equation}

Then, solve equation (\ref{6}) under the given periodic boundary conditions for the solution. This gives the admissible values of $\omega$ and therefore for $\lambda_0$. For instance if we take $p=2\pi$,  we obtain $\omega_j=2j$ with $j$ integer. Therefore, we have the following sequence of admissible seeds, labeled by the integer value $j$:

\begin{equation}\label{7}
\lambda_{0,j}=\frac{4j^2+\langle V(x)\rangle}{\langle a(x)\rangle}\,.
\end{equation}

Other two iterative methods are discussed next.

\section{Two other iterative methods.}

Along the present section, we introduce two other iterative methods based on segmentary integration. The first one, here called the {\it matrix method}, is applicable to linear systems only. This means that, if equations (\ref{1}) are not linear, another method has to be applied to find their segmentary approximate solution. In that case, we propose another kind of iterative method based in the Taylor method. The order of the method (degree of the used polynomial) depends on the desired accuracy of the segmentary solution. It is noteworthy that this iterative method, intended for non-linear systems of the form (\ref{1}) has the great advantage of being simpler to use than the usual Runge-Kutta method \cite{RK,RK1}. 

These methods are both iterative and algorithmic. Their precision will be tested in the next section. 

\subsection{Matrix method}

This first method can be applied to first order linear systems only \cite{LL}.  In general, such a linear system has the form $W'(x)=A_\lambda(x)W(x)$, where $W(x)$ is the column vector of the undetermined functions, $A_\lambda(x)$ a square matrix depending continuously on the variable $x$ and linearly on the parameter $\lambda$ and the prime denotes derivative with respect to $x$.  System (\ref{1}) can be written on this form with $W(x)=(y(x),z(x))^T$, where the superscript $T$ means transpose. Obviously, equation (\ref{3})  may also be written in the same form. 

We want to determine an approximate solution of the system $W'(x)=A_\lambda(x)W(x)$ using a segmentary procedure \cite{LL}. To implement it, we consider an initial value $W(0)=(u,v)^T$, where $u$ and $v$ are real number to be assigned. Then, divide the interval $(0,p)$ into equally spaced segments $I_{k+1}=(x_k,x_{k+1})$, with $x_k=kh$, $h=p/n$ and $k=0,1,2,\dots,n$.  On each segment $I_k$ we use the approximation $A_\lambda(x)\approxeq A_\lambda(x_k)$.  

Then, we proceed with the integration on $I_1=(0,x_1)$ of the system with constant coefficients $W'(x)=A_\lambda(0)W(x)$. Its solution gives the approximate solution on the interval $I_1$ and the value of this solution at the point $x_1$ is $W(x_1)=\exp(A_\lambda(0)h)\,W(0)$. Similarly, using $W(x_1)$ as initial condition, we integrate $W'(x)=A_\lambda(x_1)W(x)$ on $I_2=(x_1,x_2)$. Then, we repeat the process on each of the interval $I_k$, using the value $W(x_{k-1})$ as initial condition. In the last step, we obtain

\begin{equation}\label{8}
W(p)=\prod_{k=0}^{n-1} \exp(A_\lambda(x_k)h)\,W(0)\,.
\end{equation}

In order to obtain the eigenvalue $\lambda$, also called the characteristic value, we make use of one of the periodic boundary conditions $W(p)=\alpha W(0)$, $\alpha=\pm1$. Then, we arrive to the eigenvalue equation

\begin{equation}\label{9}
B_\lambda W(0)=0\,,\qquad {\rm with} \qquad B_\lambda= \prod_{k=0}^{n-1} \exp(A_\lambda(x_k)h)-\alpha I\,,
\end{equation}
where $I$ is the identity matrix. 

Consider now, $\det B_\lambda=0$.  This is an algebraic equation of order $n$, whose solutions, i.e., the values of $\lambda$ can be determined by the use of, say, Mathematica. Fix one of these values of $\lambda$; then, the eigenvalue equation in (\ref{9}) gives $W(0)=(u,v)^T$ save for a multiplicative constant, so that it is always possible to choose $u=1$.  

Then, one way to obtain $v$ and therefore the initial condition $W(0)$ for each eigenvalue $\lambda$ is the following: First of all, note that each component in (\ref{9}) must be linear on $v$. Then, for the first component in (\ref{9}), that we denote here as $(B_\lambda W(0))_1=0$, we obtain a relation $v=v(\lambda)$. Once we have determined the initial condition for a given $\lambda$, we have obtained the approximate solution for $W(x)$. 

A second operation valid to determine the solutions of the eigenvalue problem (\ref{9}), and therefore the approximate solution to $W'(x)=A_\lambda(x)W(x)$, goes in this way: From $(B_\lambda W(0))_1=0$, derive the relation $v=v(\lambda)$. We still do not know the eigenvalues $\lambda$. To determine these $\lambda$'s, let us consider the polynomial $Q(\lambda):= (B_\lambda W(0))_2$, where the subscript 2 denotes second component in the eigenvalue equation (\ref{9}). The roots of $Q(\lambda)$ give the eigenvalues, hence the values of $v$ and the segmentary eigenfunction $W(x)$. 

\subsubsection{Application via Riccati equation}

Let us consider a Schr\"odinger type equation, which can be written in the form

\begin{equation}\label{10}
y''(x)+(\lambda-V(x))y(x)=0\,,
\end{equation}
and let us define a new indeterminate as $w(x):=y'(x)/y(x)$. This is the typical substitution that transforms a Schr\"odinger equation into an inhomogeneous Riccati equation:

\begin{equation}\label{11}
w'(x)+w^2(x)= V(x)-\lambda\,,
\end{equation}
with the boundary condition $w(0)=w(p)$. Then, on each interval $I_k=(x_k,x_{k+1})$, we approximate $V(x)-\lambda$ by $V(x_k)-\lambda=V_k-\lambda$ (this defines $V_k$). After this approximation, we can obtain a recursive solution (constant on each interval $I_k$) of equation (\ref{11}), which is given on the interval $I_{k+1}$ by

\begin{equation}\label{12}
w_{k+1}=\sqrt{V_k-\lambda} \;\frac{w_k+ \sqrt{V_k-\lambda}\,\tanh(h\,\sqrt{V_k-\lambda})}{\sqrt{V_k-\lambda} +w_k \,\tanh(h\,\sqrt{V_k-\lambda})}\,,
\end{equation}
with $k=0,1,\dots,n-1$. Taking into account the boundary condition, this shows that

\begin{equation}\label{13}
w(0)-w(p)=0\,.
\end{equation}
This gives an equation on $\lambda$ whose solutions are the eigenvalues (characteristic values). Note that the dependence in $\lambda$ in equation (\ref{11}) appears in $w(p)$ only, so that this equation depends on $w(0)$, which should be fixed conveniently for each particular situation investigated. 

\subsection{Iterative Taylor method}

As we have seen, the matrix method is useful for homogeneous linear systems with variable coefficients. However, it is not applicable when the system is not linear. In this case, we propose an iterative method which is based in the Taylor expansion  \cite{HTD}, which is simpler than other usual approximations based in differential transformations like the celebrated Differential Transformation Method (DTM) \cite{DTM,chinos}. In our opinion, this implementation simplicity makes it particularly attractive and useful.

In this approach, we obtain segmentary approximate solutions by a Taylor expansion of the indeterminate functions ($y(x)$ and $z(x)$ in the case of equation (\ref{1})).  Needless to say that this requires that the indeterminate functions be differentiable up to a given order.  First of all, we choose the intervals $I_k$ as before. On each of the intervals $I_k$,  let us use the Taylor theorem in order to approximate he functions $y(x)$ and $z(x)$ in (\ref{1}) by

\begin{equation}\label{14}
y_m(x):= \sum_{j=0}^m  \frac 1{j!} \,y^{(j)}(x_k)\,(x-x_k)^j\,,  \quad z_m(x):= \sum_{j=0}^m  \frac 1{j!} \,z^{(j)}(x_k)\,(x-x_k)^j\,,
\end{equation}
where we have chosen $x_0:=0$. Here, $y^{(j)}(x)$ and $z^{(j)}(x)$ are the $j$-th derivatives of the functions $y(x)$ and $z(x)$. Their values at the points $x_k$ are to be determined via equation (\ref{1}). For $x_0:=0$, we fix some initial values $u:=y(0)$ and $v:=z(0)$, so that for the first derivative, we have

\begin{equation}\label{15}
y'(0)=f_\lambda(0,u,v)\,,\qquad z'(0)=g_\lambda(0,u,v)\,.
\end{equation}

For the second derivative, we take into consideration that

\begin{eqnarray}\label{16}
y^{(2)}(x)=\frac{\partial f_\lambda}{\partial x}+f_\lambda\,\frac{\partial f_\lambda}{\partial y} +g_\lambda\,\frac{\partial f_\lambda}{\partial z}\,, \quad z^{(2)}(x)=\frac{\partial g_\lambda}{\partial x}+f_\lambda\,\frac{\partial g_\lambda}{\partial y} +g_\lambda\,\frac{\partial g_\lambda}{\partial z}\,,
\end{eqnarray}
and proceed similarly for successive derivatives.  The initial conditions $u$ and $v$ are not yet determined. 

After (\ref{14}), the approximate solution on $I_1$ for $y(x)$ is given by

\begin{equation}\label{17}
y_m(x)=u+y'_m(0)(x-0)+\frac12\,y''(0)(x-0)^2+\dots+\frac1{m!}\,y_m^{(m)}(0)(x-0)^m\,.
\end{equation}
Same for $z(x)$ with $z_m(0)=v$. These functions defined on the first interval $I_1$ give the values $y_m(x_1)$ and $z_m(x_1)$. Following the same procedure, we use $y_m(x_1)$ and $z_m(x_1)$ as initial values for the solutions (\ref{14}) on $I_2$ and so on. At the final step, we obtain $y_m(x_n)$ and $z_m(x_n)$, which have to depend on $u$, $v$ and $\lambda$.

Next, we use boundary conditions (\ref{2}) on the approximate solutions obtained as above. Then, $u$ and $v$ have to be chosen so that the solutions obey to the required parity. Once, $u$ and $v$ have been fixed, we obtain an equation solely on $\lambda$, which determines suitable values of the eigenvalue $\lambda$.  

In the particular case in which $f_\lambda$ and $g_\lambda$ as in (\ref{1}) are linear on $y$ and $z$,  system (\ref{2}) is linear and homogeneous on $u$ and $v$.  Then, the determinant of the coefficient matrix, $\Delta$, vanishes. Under this linearity hypothesis, $\Delta=0$ is just an algebraic equation on $\lambda$. Its roots are an algebraic function of the parameters in (\ref{1}). The values for $n$ (number of intervals) and $m$ (degree of the polynomials (\ref{14})) are fixed empirically in order to obtain the desired accuracy. The use of Mathematica is now an important tool in our calculations of both Taylor coefficients and values of $\lambda$.

A discussion on the convergence of the method is given next. 

\section{On the problem of uniform convergence for the Taylor method}

We want to show that the segmentary approximate solutions converge uniformly to the exact solution for equations of the form $y'(x)=f(x,y(x))$, where $f(x,y)$ satisfy a Lipschitz condition with respect to the $y$ variable. Due to the form of system (\ref{1}), this should be sufficient in the present case. 

Assume that the number of intervals is just $n$. Then, the approximate segmentary solution is

\begin{equation}\label{55}
T_n(x):=\{t_{n,k}(x)\,,\; k=1,2,\dots,n\}\,,
\end{equation}
where $t_{n,k}(x)$ is the polynomial for the interval $I_k$.  Since we want that $T_n(x)$ be an approximation for the solution of $y'(x)=f(x,y(x))$, we should have

\begin{eqnarray}\label{56}
T'_n(x)=f(x,T_n(x))+\eta_n(x)\,, \nonumber\\[2ex]
\eta_n(x)= T'_n(x) - f(x,T_n(x))\,,
\end{eqnarray}
where $\eta_n(x)$ appears due to the discrepancy between the exact solution and the approximate solution $T_n(x)$.  An obvious integration on (\ref{56}) gives

\begin{equation}\label{57}
T_n(x)=y_0 +\int_{x_0}^x f(t,T_n(t))\,dt+\int_{x_0}^x \eta_n(t)\,dt\,,
\end{equation}
where $y_0$ is some initial condition. The point $x$ is arbitrary in the integration interval, $x\in(x_0,p]$, $p=x_n$,  so that it belongs to one of the subintervals, say $x\in I_K$. After (\ref{18}) and (\ref{56}), we can write

\begin{eqnarray}\label{58}
\int_{x_0}^x \eta_n(t)\,dt = \int_{x_0}^x [T'_n(x)-f(t,T_n(t))]\,dt  \nonumber\\[2ex]  = \sum_{k=1}^{K-1} \int_{x_k}^{x_{k+1}}[t'_{n,k+1}(t)-f(t,t_{n,k+1}(t))]\,dt  \nonumber\\[2ex] + \int_{K-1}^x [t'_{n,k+1}(t)-f(t,t_{n,k+1}(t))]\,dt\,.
\end{eqnarray}

Then, taking (\ref{14}) into account, we arrive to the following expression:

\begin{eqnarray}\label{59}
\int_{x_0}^x \eta_n(t)\,dt= \sum_{k=1}^{K-1} \left[ h_n \,a_{n,k+1} -\int_k^{x_{k+1}}  f(t,t_{n,k+1}(t))\,dt  \right]  +LT\,, 
\end{eqnarray}
where $LT$ is the last term in (\ref{58}).   Here, 

\begin{equation}\label{60}
a_{n,k+1}= \sum_{j=1}^m \frac1{j!}\,y^{(j)}(x_{k+1})\, h_n^{j-1}
\end{equation}
and $h_n$ is the length of each subinterval. Observe that the last term in (\ref{21}) is indeed similar to the other with the only difference that $h_n$ should be replaced by a number smaller or equal. 

Now after (\ref{57}), we have that

\begin{eqnarray}\label{61}
|T_{n+m}(x)-T_n(x)|\le \int_{x_0}^x |f(t,T_{n+m}(t))-f(t,T_n(t))|\,dt+\varepsilon_{nm}\,,
\end{eqnarray}
with

\begin{equation}\label{62}
\varepsilon_{n,m}=\left|\int_{x_0}^x [\eta_{n+m}(t)-\eta_n(t)]\,dt \right|\,.
\end{equation}

Let us use the Lipschitz condition in (\ref{61}). This gives:

\begin{equation}\label{63}
|T_{n+m}(x)-T_n(x)|\le R \int_{x_0}^x  |T_{n+m}(t)-T_n(t)|\,dt+ \varepsilon_{nm}\,,
\end{equation}
where $R$ is a constant. Each of the functions in the sequence $\{T_n(x)\}$ is bounded, so that after (\ref{26}) we can obtain

\begin{equation}\label{64}
\sup_{x\in[x_0,p]} |T_{n+m}(x)-T_n(x)|\le \frac{\max \varepsilon_{n,m}}{|1-R(x-x_0)|}\,.
\end{equation}

It is important to mention that $T_{n+m}(x)$ and $T_n(x)$ correspond to two different partitions of the integration interval. In these two different cases the partition has $n+m$ and $n$ subintervals, respectively. We shall denote by $\mathcal P_n$ to the partition with $n$ intervals. Without loss of generality, we may assume that $n=2^p$, where $p$ is a natural number, so that $\mathcal P_m$ with $m>n$ is always a refinement of $\mathcal P_n$. 

In order to study the term $\varepsilon_{n+m}$, let us go back to (\ref{59}) and consider

\begin{eqnarray}\label{65}
\int_{x_0}^x [\eta_{n+m}(t)-\eta_n(t)]\,dt = \sum_{k=1}^{K_{(m+n)}-1}  h_{n+m}\,a_{n+m,k} - \sum_{k=1}^{K_{(m)}-1}  h_{n}\,a_{n,k} \nonumber\\[2ex] \sum_{k=1}^{K_{(m)}-1} \int_{x_k}^{x_{k+1}} f(t,t_{n,k+1}(t)) \,dt  - \sum_{k=1}^{K_{(m+n)}-1} \int_{x_k}^{x_{k+1}} f(t,t_{n+m,k+1}(t)) \,dt \nonumber\\[2ex]+LT_{(n+m)}-LT_{(n)} \,.
\end{eqnarray}

The notation we have used in (\ref{28}) the symbol $K_{(p)}$ in the upper limit of the sum means that we have used the partition $\mathcal P_p$. The meaning of the two last terms should be obvious. Since we have chosen $\mathcal P_{n+m}$ to be a refinement of the partition $\mathcal P_n$, we may stand on the former. Then, using the definition (\ref{25}), we have the following inequality

\begin{eqnarray}\label{66}
\max\varepsilon_{n,m}\le \left[\sum_{k=1}^{K_{(m+n)}}  \max |a_{n+m,k} -  a_{n,k}|  \right]\, h_{n+m} \nonumber\\[2ex] +
\sum_{k=1}^{K_{(m+n)}} \int_{x_k}^{x_{k+1}} |f(t,t_{n,k+1}(t))- f(t,t_{n+m,k+1}(t))|\,dt\,.
\end{eqnarray}

In (\ref{29}) we have included the two last terms in (\ref{28}). 

Due to their definition, the coefficients $a_{p,k}$ are uniformly bounded for any partition $\mathcal P_p$. The polynomials $t_{n,k}(t)$ and $t_{n+m,k}(t)$ have the same degree so that on each of the subintervals $I_{n+m}$, one has

\begin{equation}\label{}
|t_{n,k}(t)-t_{n+m,k}(t)| \le \alpha_{n,m,k}\, h_{n+m}\,,
\end{equation}
with $\alpha_{n,m,k}\longmapsto 0$ when $n,m\longmapsto\infty$.  In addition, the functions $T_n(x)$ and $T_{n+m}(x)$ are continuous on the same finite interval, so that the $\alpha_{n,m,k}$ are uniformly bounded by some $\alpha$.  Using the Lipschitz condition for the function $f(x,y)$ with constant $R$ and (\ref{30}) and taking into account that $h_r=p/r$, the term with the integral in (\ref{66}) is bounded by

\begin{eqnarray}\label{}
\sum_{k=1}^{K_{(m+n)}} \int_{x_k}^{x_{k+1}} |f(t,t_{n,k+1}(t))- f(t,t_{n+m,k+1}(t))|\,dt  \nonumber\\[2ex] \le \sum_{k=1}^{K_{(m+n)}} \alpha\, R\,  h_{n+m}^2 =\alpha\,(n+m)\,R\,h_{n+m}^2= \alpha R\,\frac{p^2}{n+m} \longmapsto 0\,,
\end{eqnarray}
as $n+m\longmapsto\infty$. Thus, the second term in (\ref{29}) goes to zero. 

Concerning the first term in (\ref{66}). The derivatives $y^{(j)}(x)$, $j=1,2,\dots,s$, are all bounded on the integration interval, because of their continuity on a compact interval. Then taking into account the explicit form of the $a_{n,k}$ given in (\ref{23}), it is not difficult to show that this first term in (\ref{}) also goes to zero as $n+m\longmapsto\infty$. In consequence,

\begin{equation}\label{69}
\lim_{n+m\mapsto \infty} \max \varepsilon_{n,m}=0\,.
\end{equation}

Let us go back to (\ref{64}) and note that this inequality does not guarantee the uniform convergence of the sequence $T_n(x)$ if there exists an $x\in[x_0,p]$ such that the denominator in the right hand side of (\ref{64}) vanishes. However, uniform convergence is assured if the interval width is smaller than $R^{-1}$, since in this case no such an $x$ may exist. 

If the width of the interval $[x_0,p]$ were larger than $R^{-1}$, in order to ensure uniform convergence, let us choose $p_1:< x_0+1/R$ and apply the procedure described on Section 3.2 to this interval.  Then, repeat the method on the interval $[p_1,p_2]$ with $p_2<p_1+1/R$ and so on. 

We have shown that the sequence of  approximate solutions $\{T_n(x)\}$ converges uniformly to a function $T(x)$. Then, using (\ref{}), the properties of the functions involved in this relation and the Lebesgue theorem, we conclude that

\begin{equation}\label{70}
T(x)=y_0+\int_{x_0}^x f(t,T(t))\,dt\,,
\end{equation}
so that $T(x)$ is the solution of $y'(x)=f(x,y(x))$ with initial value $y(x_0)=y_0$. 

\section{Numerical comparison of the methods}

As a laboratory to compare the efficiency and numerical accuracy of our method, we use here the Mathieu equation \cite{AS}. In addition, we are going to apply our methods to some situations with physical relevancy, in particular to graphene nanotubes. If we take as reference that particular case, we obtain equations with periodical coefficients and one possible approximation gives the Mathieu equation. It has the following form:

\begin{equation}\label{18}
y''(x)+(r-2q\cos(2x))y(x)=0\,.
\end{equation}

In (\ref{18}), $q$ is a given parameter, while $r$ is the characteristic value or eigenvalue which should be determined. 
There are four series of periodic solutions of (\ref{18}), each one labeled with a discrete series of characteristic values \cite{FAR}. Here we choose even periodic solutions on the interval $[0,2\pi]$, which are usually written in the form \cite{BC,GGL}: 

\begin{equation}\label{19}
y(x,q,r_{2m+1}(q))=\sum_{k=0}^\infty A_k\,\cos(2k+1)x\,,\qquad  m=0,1,2,\dots\,.
\end{equation}

If we take for instance $q=1$, the three first characteristic values are $r_1=1.85911$, $r_3= 9.07837$ and $r_5= 25.0209$. See \cite{AS}. Once we have determined these three first characteristic values, let us define as customary the percentage relative error as

\begin{equation}\label{20}
\epsilon_r\,\%:= 100\,\left| \frac{r_{num}-r_{exact}}{r_{exact}}   \right|\,,
\end{equation}
where $r_{num}$ is the characteristic value obtained by a numerical procedure in contrast with its exact value $r_{exact}$. We have determined the cpu time, $t_{cpu}$ using the software Mathematica 9.0 and the hardware AMD Athol (tm) II X2 250 Processor with 4 GB RAM.  All methods under our consideration require an initial value of the characteristic value, which we denote as $r^*$. This will be the seed for the SAM iterations or as the initial value for the calculation of the roots in the matrix method. Since the cpu time $t_{cpu}$ depends on the chosen seed, in order to compare the different cases we have taken $r^*:=1.05 r_{exact}$.

Thus, the numerical values we have obtained are the following:

\medskip
1.- Successive Approximation Method. In the next table, we show the results obtained via SAM for the three first characteristics values and the percentage relative error in terms of the number $j$ of iterations:

\bigskip

$
\begin{array}
[c]{ccc}
j & r_{1} & \epsilon_{r}\,\%\\
1 & 1.90284 & 2.35\\
2 & 1.86813 & 0.48\\
3 & 1.85947 & 0.02\\
4 & 1.85911 & 10^{-5}\\
5 & 1.85911 & 410^{-5}
\end{array}
$ \ \ \ \ \ \ $
\begin{array}
[c]{ccc}
j & r_{3} & \epsilon_{r}\,\%\\
1 & 9.02009 & 0.64\\
2 & 9.07842 & 6.10^{-4}\\
3 & 9.07837 & 5.10^{-7}\\
4 & 9.07837 & 5.10^{-7}\\
5 & 9.07837 & 5.10^{-7}
\end{array}
$ \ \ \ \ \ \ \ \ $
\begin{array}
[c]{ccc}
j & r_{5} & \epsilon_{r}\,\%\\
1 & 25.4707 & 1.79\\
2 & 25.0463 & 0.10\\
3 & 25.0209 & 5.10^{-5}\\
4 & 25.0209 & 8.10^{-8}\\
5 & 25.0209 & 8.10^{-8}
\end{array}
$

$t_{cpu}=2.28s$ \ \ \ \ \ \ \ \ \ \ \ \ \ \ \ \ \ \ \ \ $t_{cpu}=3.42$ $s$
\ \ \ \ \ \ \ \ \ \ \ \ \ \ \ \ \ \ \ $t_{cpu}=5.05s$

\bigskip

Here, cpu times grow due to the increase in the oscillations on $y$, which affects to the numerical evaluation of  integrations. 

\medskip
2.- Matrix method (MM). Using this method, we show the results obtained, when we divide the interval $(0,2\pi)$ into 10 and 12 subintervals, respectively (here  and also in the following tables, $n$ denotes the number of intervals):

\bigskip

$
\begin{array}
[c]{ccc}
k & r_{k} & \epsilon_{r}\,\%\\
1 & 1.81017 & 2.6\\
3 & 9.0651 & 0.14\\
5 & 25.0597 & 0.15
\end{array}
$ \ \ \ \ \ \ \ \ \ $
\begin{array}
[c]{ccc}
k & r_{k} & \epsilon_{r}\,\%\\
1 & 1.82598 & 1.8\\
3 & 9.0707 & 9.10^{-2}\\
5 & 25.207 & 0.8
\end{array}
$

\smallskip
$t_{cpu}=15.9$ $s\ \ n=10$ \ \ \ \ \ \ $t_{cpu}=76.1$ $s\ \ n=12$

\bigskip

Now the $t_{cpu}$ are too high due to the calculation of the roots. In order to improve these results, we rewrite the Mathieu equation in the Riccati form. Since we are looking for evan solutions, we take $w(0)=1$. Now we obtain a much better result as can be seen in the next table:

\bigskip

$
\begin{array}
[c]{ccc}
k & r_{k} & \epsilon_{r}\,\%\\[2ex]
1 & 2.23995 & 20.5\\
3 & 8.92347 & 1.70\\
5 & 24.8458 & 0.70
\end{array}
$ \ \ \ \ \ \ \ \ \ $
\begin{array}
[c]{ccc}
k & r_{k} & \epsilon_{r}\,\%\\[2ex]
1 & 1.85539 & 0.20\\
3 & 9.07428 & 5.10^{-2}\\
5 & 25.0184 & 10^{-3}
\end{array}
$ \ \ \ \ \ \ \ \ \ \ \ $
\begin{array}
[c]{ccc}
k & r_{k} & \epsilon_{r}\,\%\\[2ex]
1 & 1.8584 & 4.10^{-2}\\
3 & 9.07768 & 7.10^{-3}\\
5 & 25.0205 & 1.10^{-3}
\end{array}
$

\smallskip
$t_{cpu}=0.0654\, s$, $ n=10$ \ \ \ \ $t_{cpu}=1.64\,s$, $n=50$
\ \ \ \ \ \ \ $t_{cpu}=7.21\,s$, $n=100$

\bigskip

3.- Finally, we show the results obtained by the iterative Taylor method. Let us start with second order:

\bigskip

$
\begin{array}
[c]{ccc}
k & r_{k} & \epsilon_{r}\,\%\\[2ex]
1 & 1.75409 & 5.64\\
3 & 3.19332 & 65.\\
5 & 0.198295 & 99.
\end{array}
$ \ \ \ \ \ \ \ \ \ $
\begin{array}
[c]{ccc}
k & r_{k} & \epsilon_{r}\,\%\\[2ex]
1 & 1.84606 & 0.70\\
3 & 8.679 & 4.5\\
5 & 22.5466 & 10.
\end{array}
$ \ \ \ \ \ \ \ \ \ \ \ $
\begin{array}
[c]{ccc}
k & r_{k} & \epsilon_{r}\,\%\\[2ex]
1 & 1.85575 & 0.18\\
3 & 8.96382 & 1.2\\
5 & 24.4015 & 2.5
\end{array}
$

\smallskip
$t_{cpu}=0.17$ $s\ \ n=10$ \ \ \ \ \ \ $t_{cpu}=1.28$ $s\ \ n=50$
\ \ \ \ \ \ \ $t_{cpu}=4.64$ $s\ \ n=100$

\bigskip

Note that for second order, times $t_{cpu}$ are reasonable. However, we do not consider the accuracy as acceptable and therefore, we make the same numerical analysis using fourth order Taylor. We obtain:

\bigskip

$
\begin{array}
[c]{ccc}
k & r_{k} & \epsilon_{r}\,\%\\[2ex]
1 & 1.85725 & 10^{-1}\\
3 & 8.6328 & 4.9\\
5 & 20.4753 & 18.
\end{array}
$ \ \ \ \ \ \ \ \ \ $
\begin{array}
[c]{ccc}
k & r_{k} & \epsilon_{r}\,\%\\[2ex]
1 & 1.85917 & 3.10^{-3}\\
3 & 9.08136 & 3.10^{-2}\\
5 & 25.0932 & 0.29
\end{array}
$ \ \ \ \ \ \ \ \ \ \ \ $
\begin{array}
[c]{ccc}
k & r_{k} & \epsilon_{r}\,\%\\[2ex]
1 & 1.85911 & 2.10^{-4}\\
3 & 9.07857 & 2.10^{-3}\\
5 & 25.0275 & 3.10^{-2}
\end{array}
$

\smallskip
$t_{cpu}=0.21$ $s\ \ n=10$ \ \ \ \ \ \ $t_{cpu}=2.82$ $s\ \ n=50$
\ \ \ \ \ \ \ $t_{cpu}=12.5$ $s\ \ n=100$

\bigskip

From the precedent tables, we observe the following: 

i.) The matrix method for Riccati has the same level of accuracy like the Taylor method with $n=100$. 

ii.) Contrarily as it happens with the Taylor method, the matrix method in Riccati reduces the error when computing larger eigenvalues. 

iii.) The SAM is more efficient for  the Mathieu equation: we just need four iterations to obtain the eigenvalues $r_1$, $r_3$ and $r_5$ with better accuracy than in the other case and with  $t_{cpu}\approxeq 10$ seconds. 

\medskip

Finally, in order to close our discussion, we perform the same calculations using the classical finite difference method, which approaches the second derivative of $y(x)$ evaluated at $x_k$ by its discrete derivative $h^{-2}(y_{k-1}-2y_k+y_{k+1})$:

\bigskip

$
\begin{array}
[c]{ccc}
k & r_{k} & \epsilon_{r}\,\%\\[2ex]
1 & 1.74669 & 6.1\\
3 & 9.89793 & 9.1\\
5 & 12.6158 & 49.6.
\end{array}
$ \ \ \ \ \ \ \ \ \ $
\begin{array}
[c]{ccc}
k & r_{k} & \epsilon_{r}\,\%\\[2ex]
1 & 1.85155 & .41\\
3 & 8.94667 & 1.5\\
5 & 21.5864 & 13.8
\end{array}
$ \ \ \ \ \ \ \ \ \ $
\begin{array}
[c]{ccc}
k & r_{k} & \epsilon_{r}\,\%\\[2ex]
1 & 1.85618 & 0.16\\
3 & 9.02127 & 0.63\\
5 & 24.826 & 0.78
\end{array}
$

\smallskip
$t_{cpu}<0.1$ s, $n=10$ \ \ \ \ \ \ \ $t_{cpu}=0.17$ s, $n=50$
\ \ \ \ \ \ \ $t_{cpu}=0.44$ s, $n=100$

\bigskip
$ 
\begin{array}
[c]{ccc}
k & r_{k} & \epsilon_{r}\,\%\\[2ex]
1 & 1.85889 & 1.10^{-2}\\
3 & 9.09425 & 0.18\\
5 & 25.0052 & 6.10^{-2}
\end{array}
$

\smallskip
 $t_{cpu}=32.6$ s,  $n=1000$.

\bigskip

One should note that in this case, we need much higher values of $n$ in order to obtain a similar precision than for the previous methods, along with a non desirable significative  increase of cpu times. 

\section{Application: Carbon nanotubes in a transverse magnetic field depending on elliptic functions.}

In a recent article  \cite{JKN}, Jakubsk\'y et al. have discussed an exactly solvable model in which a transverse magnetic field interacts with the electrons in a single-wall nanotube. After the choice of the interaction in \cite{JKN}, the exact solvability was achieved by fixing one parameter equal to zero. For non-zero values of this parameter, approximate methods should be applied. We want to undergo this task in the present Section. 

The nanotube under consideration consists in a one atom thick layer of carbon having the form of a cylinder of infinite heigh and radius $\rho_0$. On the nanotube, a magnetic field $\bf B$ acts with a vector potential ${\bf A}=A_\phi{\bf n}_\phi+A_z{\bf n_z}$, where ${\bf n}_\phi$ and ${\bf n_z}$ are the unit vectors tangent to the circumference and to the longitudinal direction. In \cite{JKN}, the component $A_\phi$ was assumed to be constant, while $A_z$ is a function of the angle $\phi$ alone, $A_z\equiv A_z(\phi)$. Along the present article, we are keeping the same assumptions, were $A_\phi$ and $A_z(\phi)$ are given data. 

The behavior of a single electron on the nanotube is governed by a Dirac type equation of the form \cite{JKN}:

\begin{equation}\label{21}
\left[ \sigma_1\frac{i}{\rho_0}\;\partial_\phi-\sigma_2\left(i\partial_z+\frac{q}{c\hbar}\;A_z(\phi)  \right) \right]\widetilde\Psi(z,\phi)=\epsilon\widetilde\Psi(z,\phi)\,,
\end{equation}
where $\sigma_1$ and $\sigma_2$ are the $2\times 2$ Pauli matrices \cite{COHEN}, $q$, $c$ and $\hbar$ are the electron charge, the speed of light in the vacuum and the Planck constant divided by $2\pi$ respectively. In (\ref{21}),  $\epsilon=E/(v_F\hbar)$, where $E$ is the energy and  $v_F$ is the Fermi velocity in the graphene whose value is $v_F\approx 10^6$ m/s. 

Note that equation (\ref{21}) is linear with indeterminates $\epsilon$ and $\widetilde\Psi(z,\phi)$.  Thus, we should look for factorizable solutions on both variables $z$ and $\phi$. Since (\ref{21}) is  a Dirac equation, it has two components. Its solution in terms of the indeterminate $\widetilde\Psi(z,\phi)$ should have the form

\begin{equation}\label{22}
\widetilde\Psi(z,\phi)=e^{ik_zz}\,\Psi(\phi)\,,\qquad \Psi(\phi)=\left(\begin{array}{c} \psi_+(\phi)\\[2ex] \psi_-(\phi)  \end{array} \right)\,.
\end{equation}

 After (\ref{22}), equation ({\ref{21}) depends on the variable $\phi$ alone so that (\ref{21}) is transformed into

\begin{equation}\label{23}
\left( i\sigma_1\partial_\phi+\left( \rho_0k_z-\frac{2\pi\rho_0}{\Phi_0}\;A_z(\phi)  \right) \sigma_2\right)\Psi(\phi)=\rho_0\epsilon\Psi(\phi)\,,  \quad \Phi_0:=\frac{2\pi c\hbar}q\,.
\end{equation}

This is an equation with two components, which give a system like (\ref{1}) with $y\equiv\psi_+$ and $z\equiv\psi_-$. This may be replaced by a second order equation like   (\ref{3}). We shall proceed with this manipulation later. 

The solvability of the model  depends on the choice of the vector potential $A_z(\phi)$ as well as the value of the parameter $k_z$ \cite{JKN}. This solvability is assured taken $k_z=0$ and 

\begin{equation}\label{24}
A_z(\phi)=\rho_0B_0 a(\phi,k)\,,
\end{equation}
where,

\begin{equation}\label{25}
a(\phi,k)=(1+k')\frac{sn[ (\phi+\pi/2)K/\pi]\,cn[(\phi+\pi/2)K/\pi ]}{dn[(\phi+\pi/2)K/\pi ]}\,.
\end{equation}

Here, $sn(x,k)$, $cn(x,k)$ and $dn(x,k)$ are the Jacobi elliptic functions and $K$ is a function of $k$ given by the following elliptic integral \cite{AS}:

\begin{equation}\label{26}
K(k):=\int_0^{\frac\pi2}\frac{dt}{1-k^2\sin^2t}\,.
\end{equation}

The parameter $k\in[0,1]$ is called the modular parameter. Sometimes one also uses $k':= \sqrt{1-k^2}$. Note that (\ref{25}) and the properties of the elliptic functions imply that $\lim_{k\mapsto 0}a(\phi,k)=\cos \phi$.  

As mentioned earlier, equation (\ref{23}) is exactly solvable for the particular case  $k_z=0$.  The given solution shows  two energy bands with positive energy with one gap between them. The first energy band is finite while the second one is unbounded \cite{JKN}.  The purpose of this section is to analyze some aspects on the behavior of the energy bands when $k_z\ne 0$ using the approximate methods described in Section 3. 

\subsection{Discussion on the model}

The value of $k$ in (\ref{26}) varies from 0 to 1 and is {\it fixed} in the model proposed in \cite{JKN}.  In order to make the calculations with Mathematica easier, let us use instead $m:=k^2$.  Then in (\ref{25}), $(k')^2=1-m$. Let us also define (see (\ref{25})) the variable:

\begin{equation}\label{27}
x:= \frac{(\phi+\pi/2)K(m)}{\pi}\,,
\end{equation}
where $m$ is fixed so that $x$ depends solely on the angle $\phi$.  Also, for constant $m$ the Jacobi functions depend only on te variable $x$, but in any case, we shall write $sn(x,m)$, $cn(x,m)$ and $dn(x,m)$ for these Jacobi functions. 

Then, we can write the two components matrix equation (\ref{23}) as

\begin{equation}\label{28}
H\Psi(x)=\frac{\pi\rho_0}{K(m)}\,\epsilon \,\Psi(x)\,, \qquad \Psi(x)=\left(\begin{array}{c}    \varphi_+(x) \\[2ex]  \varphi_-(x)\end{array}    \right)\,,
\end{equation}
where $H$ has the following form:

\begin{equation}\label{29}
 H=\left( \begin{array}{cc}  0  &   iA^+ \\[2ex]  -iA^-  &  0         \end{array}        \right)\,, \qquad {\rm with}  \qquad A^\pm=\pm \partial_x+W(x)\,.
\end{equation}

Here, $\partial_x$ denotes derivative with respect to $x$ and 

\begin{equation}\label{30}
W(x)= -\frac{\pi\rho_0}{K(m)}\,k_z+W_0(x)\,,\qquad W_0(x)= 2m \,\frac{sn(x,m)\,cn(x,m)}{dn(x,m)}\,. 
\end{equation}

It is convenient to multiply (\ref{28}) by $H$. It produces the effect of decoupling the system. This operation transforms (\ref{28}) into

\begin{equation}\label{31}
H^2\Psi(x)= \left(\begin{array}{cc}  A^+A^-  & 0 \\[2ex]  0  &  A^-A^+   \end{array}    \right) \left(\begin{array}{c}    \varphi_+(x) \\[2ex]  \varphi_-(x)\end{array}    \right)  = \left( \frac{\pi\rho_0}{K(m)}  \right)^2 \,\epsilon^2\,\Psi(x)\,.
\end{equation}

Next, let us define 

\begin{equation}\label{32}
H_+:= A^+A^-\,,\qquad H_-:= A^-A^+\,,
\end{equation}
so that

\begin{equation}\label{33}
H^2=\left(\begin{array}{cc}  H_+  & 0\\[2ex]  0  & H_-   \end{array}   \right)\,, \qquad H_\pm= -\partial_x^2+V_\pm(x)
\end{equation}
with

\begin{equation}\label{34}
V_\pm(x):= W^2(x)\pm W'(x)\,.
\end{equation}

At this point, it is in order to comment that the chosen potential $a(\phi,k)$ given in (\ref{25}) is nothing else than the superpotential for the potential $V(\phi,k)=sn^2(\phi,k)$, which is, save for a constant, the well known Lam\'e potential \cite{CKS}. 

From ({\ref{31}) we readily obtain two second order separate equations, one for $\varphi_+(x)$ and the other for $\varphi_-(x)$. These are eigenvalue equations for $H_+$ and $H_-$, respectively, which are equally valid to obtain approximate values for the eigenvalues $\epsilon$, which  is the objective of the present analysis. To this end, let us use the eigenvalue equation for $H_-$:

\begin{equation}\label{35}
\partial_x^2\varphi_-(x)+\left( \left( \frac{\pi\rho_0}{K(m)}   \right)^2\,\epsilon^2 -V_-(x)   \right)\varphi_-(x)=0\,.
\end{equation}

We are studying properties of nanotubes in terms of the angle $\phi$. Therefore, functions depending on $\phi$ have to show periodicity properties in terms of this angle. According to \cite{JKN}, these solutions should be either periodic or anti-periodic. This periodicity properties have to be inherited by the functions $\varphi_\pm(x)$.  This comes from (\ref{28}) and (\ref{29}), which imply that $A^-\varphi(x)=s\varphi_+(x)$ and $A^+\varphi_+(x)=-s\varphi_-(x)$ with $s=i\pi\rho_0\epsilon/K(m)$. From the former of these equations and the boundary conditions for $\varphi_-(x)$, one gets $\varphi_+(0)=\pm\varphi_+(P)$. From the second and $A^+=\partial_x+W(x)$ and the previous boundary conditions, we get $\varphi'_+(0)=\pm\varphi'_+(P)$.
After (\ref{27}), the period is $P=2K(m)$. In particular, solutions of (\ref{35}) satisfy either periodicity or anti-periodicity  properties:

\begin{equation}\label{36}
\varphi_-(x)=\alpha \varphi_-(x+P)\,,\qquad   \varphi'_-(x)=\alpha \varphi'_-(x+P)\,,
\end{equation}
where $\alpha=1$ and $\alpha=-1$, respectively.  

Changes of scale usually help in simplifying calculations. In particular, we propose the following:

\begin{equation}\label{37}
x/P\longrightarrow x\,, \qquad 2\pi\rho_0\epsilon \longrightarrow \epsilon\,, \qquad P^2V_\pm \longrightarrow V_\pm\,.
\end{equation}

Under this change of scale along a unit system such that $p=1$ and $\pi\rho_0=1$, equation (\ref{35}) transforms into

\begin{equation}\label{38}
\partial_x^2\varphi_-(x)+(\epsilon^2-V_-(Px))\varphi_-(x)=0\,.
\end{equation}

With these new units, we can write $V_-(x)$ as:

\begin{eqnarray}\label{39}
V_-(x) =4k_z\left( k_z-m\,\frac{sn(2K(m)x,m)\,cn(2K(m)x,m)}{dn(2K(m)x,m)}\right)\nonumber\\[2ex]  + 4mK^2(m) (2sn^2(2K(m)x,m)-1) \,.
\end{eqnarray}

Note that for $k_z=0$, (\ref{39}) is essentially the Lam\'e potential. An Schr\"odinger equation of the type $-d^2\phi(x)/dx^2+V(x)\phi(x)=\epsilon^2\phi(x)$, where $V(x)$ is the Lam\'e potential  has analytic solution in terms of the elliptic Jacobi functions \cite{CKS}.

Let us assume that $m<<1$, for simplicity. At first order, the potential $V_-(x)$ looks like

\begin{equation}\label{40}
V_-(x)= 2k_z(k_z-2m\pi \sin(2\pi x))-m\pi^2\cos(2\pi x)\,,
\end{equation}
so that (\ref{38}) becomes a Hill equation.  Furthermore, if we choose $k_z=0$, we obtain the Mathieu equation. In Figure 1, we compare the exact and approximate potentials for small values of $m$ and $k_z=1$. We see that in the range of chosen values $m\le 0.20$, both exact and approximate potentials are quite similar.

\subsection{Some approximations}

Let us consider the following equation 

\begin{equation}\label{41}
y''(x) +(a-f(x))y(x)=0\,,
\end{equation}
where $a$ is a fixed real number, $\alpha\le x\le \beta$ and $f(x)$ continuous on $(\alpha,\beta)$. Equation (\ref{41}) can be approximated as

\begin{equation}\label{42}
\zeta''(x)+(a-\langle f(x)\rangle)\zeta(x)=0\,,
\end{equation}
where the meaning of the average $\langle f(x)\rangle$ is as in (\ref{6}). 

In order to analyze the error in the approximation, let us consider the number $x^*$ such that $f(x^*)=\langle f(x)\rangle$. We know that $x^*\in(\alpha,\beta)$ due to the mean value theorem.  Then, let us find the Taylor expansion of the solutions of (\ref{41}) and (\ref{42}) on a neighborhood of $x^*$. They are respectively: 

\begin{eqnarray}\label{43}
y(x)=y(x^*)+ y'(x^*)(x-x^*)-\frac12 (a-f(x^*)) y(x^*)(x-x^*)^2 \nonumber\\[2ex]
-\frac16 (a-f(x^*))y(x^*)(x-x^*)^3-\frac16 f'(x^*)(x-x^*)^3+\dots
\end{eqnarray}
and

\begin{eqnarray}\label{44}
z(x)=z(x^*)+ z'(x^*)(x-x^*) -\frac12 (a-\langle f(x)\rangle) z(x^*) (x-x^*)^2 \nonumber\\[2ex]
-\frac16 (a-\langle f(x)\rangle) z(x^*) (x-x^*)^3+\dots\,.
\end{eqnarray}

We see that (\ref{43}) and (\ref{44}) coincide up to second order. If we denote by $h$ the minimal radius of convergence of both series, we have that $h<\beta-\alpha$ (note that in the case of the Graphene, $\beta-\alpha=1$). Then, the error in the approximation is of the order of $h^3$.  

Thus, equation (\ref{38}) can be approximated by

\begin{equation}\label{45}
\partial^2_x \varphi_{app}(x) +(\epsilon^2 -\langle V_-(Px)\rangle)\,\varphi_{app}(x)=0\,. 
\end{equation}

The function $\varphi_{app}(x)$ should approximate $\varphi_-(x)$. Then, the solution of the Sturm-Liouville problem to (\ref{45}) with either periodic and anti periodic boundary conditions on $\varphi_{app}(x)$ is trivial. When the boundary conditions are periodic, the energy values (eigenvalues) are

\begin{equation}\label{46}
\epsilon_n^2 \approxeq (2n)^2\pi^2+ \langle V_-(Px)\rangle\,, \qquad n=1,2,3,\dots
\end{equation}

When the boundary conditions are anti periodic, we get the following energy values:

\begin{equation}\label{47}
\epsilon_n^2 \approxeq (2n+1)^2\pi^2+ \langle V_-(Px)\rangle\,, \qquad n=1,2,3,\dots
\end{equation}

It is necessary to recall that in equation (\ref{38}) the values of $\epsilon$ were rescaled. In order to obtain the true values we have to go back to the original scale. This approximation, although simple is a good one as we shall see along the next subsection.

\subsection{Some numerical results.}

Before this section, we have introduced three approximative methods that we have used on equation (\ref{38}) through a large number of numerical experiments. Our conclusion is that the most efficient of all the three is the iterative Taylor method. 

It is important to mention that the eigenvalues $\epsilon^2$, corresponding to the solution of (\ref{38}), are customarily ordered as follows:

\begin{equation}\label{48}
\epsilon^2_0< \epsilon^2_1 \le \epsilon^2_{1'}<\epsilon^2_{2} \le \epsilon^2_{2'}< \epsilon^2_3 \le \epsilon^2_{3'} < \epsilon^2_4 \dots
\end{equation}

The eigenvalues with even subindices,

\begin{equation}\label{49}
\epsilon^2_0 < \epsilon^2_2\le \epsilon^2_{2'}<\epsilon^2_4,\,,\qquad {\rm etc}\,,
\end{equation}
correspond to eigenvalues with periodic eigenfunctions. The eigenvalues with odd subindices,

\begin{equation}\label{50}
\epsilon^2_1 < \epsilon^2_3\le \epsilon^2_{3'}<\epsilon^2_5,\,,\qquad {\rm etc}\,,
\end{equation}
correspond to anti-periodic solutions.  Compare to (\ref{46}) and (\ref{47}). The forbidden energy bands are:

\begin{equation}\label{51}
(\epsilon^2_1,\epsilon^2_{1'}), \quad (\epsilon^2_2,\epsilon^2_{2'})\,,\qquad {\rm etc}\,,
\end{equation}
so that when $\epsilon^2_k=\epsilon^2_{k'}$ the corresponding forbidden band disappears. On the other hand, the permitted bands are:

\begin{equation}\label{52}
[\epsilon^2_0,\epsilon^2_1]\,, \qquad [\epsilon^2_{1'},\epsilon^2_2]\,, \qquad {\rm etc}\,.
\end{equation}

For the Lam\'e equation ($k_z=0$), we know the following eigenvalues:

\begin{equation}\label{53}
\epsilon^2_0=0\,,\qquad \epsilon^2_1=1-m\,,\qquad \epsilon^2_{1'}=1\,.
\end{equation}

Their respective eigenfunctions are 

\begin{equation}\label{54}
\psi_0(x)=dn \,x\,,\qquad \psi_1(x)=cn \,x\,,\qquad \psi_{1'}(x)= sn\,x\,,
\end{equation}
which shows that there exist two allowed and one forbidden bands \cite{JKN}. When $k_z\ne 0$ there is no analytic solutions and therefore allowed and forbidden bands are unknown.

Let us go back to equation (\ref{38}). When we used the iterative Taylor method, we have chosen a number of integration intervals $n=100$ for the Lam\'e equation. As we noted earlier, this equation has  three exact eigenvalues to compare with. For the cases in which no exact solution is available, we have compared results with $n=100$ to results with $n=200$. We have obtained a relative variation of order less than $5\cdot 10^{-2}\,\%$. 

In Figure 2, we split $\epsilon^2_1$ and $\epsilon^2_{1'}$ for increasing values of $m$, starting from $m=0$. Two curves  that have smaller energy at $m=0$ correspond to $k_z=0$, i.e., the situation for which the exact solution is known. Here, the maximal error between the exact and numerical values is of the order of $5\cdot 10^{-5}\,\%$.  For the other two with higher energy at $m=0$, we have chosen $k_z=1$, so that no exact solution is known. 

Figure 3 represents the variation of the numerical values of $\epsilon_1^2$ and $\epsilon_{1'}^2$ with $|k_z|$, for a fixed value of $m$ that we have fixed here as $m=0.5$. 

Going on with the same procedure, our numerical results show that the splitting of the levels other than the lower one is really small. To complete our analysis, let us go back to equations (\ref{46}) and (\ref{47}). Within the range $0<|k_z|<3$ and $0<m<0.7$, our results show  a maximal relative difference, $|100(\epsilon^2_{num}-\epsilon^2_{approx})/\epsilon^2_{num}|$, where $\epsilon^2_{num}$ and $\epsilon^2_{approx}$ are the values obtained numerically and from the approximations given by formulas (\ref{46}) and (\ref{47}), respectively, of the order less than $0.2\,\%$. This shows again a quite small splitting. Note this naive approximation is not really good for the ground level. In fact, $(\epsilon^2_1+\epsilon^2_{1'})/2$ can be approximated by $\epsilon^2_{approx}$ with an error smaller than $1\%$. 

Finally, we have noted that the spectrum is very little sensitive to variations on $m$ and $k_z$. 

\section{Concluding remarks}

We have developed three different iterative methods in order to determine the approximate eigenvalues of a Sturm-Lioville system, which may have applications to physics. In order to check their accuracy and applicability, we have applied these methods to the determination of eigenvalues (here called characteristic values) of the Mathieu equation. They have been studied by using another strategies, so that we have a basis for comparison. We have observed that our  successive approximation method (SAM) gives good results with low iterations and low cpu times.

In the case of the Ricccati equation, that has a particular interest in physics because it comes from a transformation on a Schr\"odinger type equation. The SAM method shows a high level of accuracy as compared to the usual Taylor method, when the interval of integration has been divided in $n=100$ subintervals. 

In the last section, we have applied our methods to calculate the eigenvalues in a model of graphene nanotubes, with either periodic or anti periodic boundary conditions. Except for a limit value on a parameter, there is not an analytic method to obtain the eigenvalues (and in that limiting case, only three eigenvalues have been determined exactly). Here, our numerical experiments have shown that our iterative Taylor method is the most efficient of the three proposed. Our result permit an evaluation of the energy levels, although they do not detect a clear removal of the degeneracy (contrary to what we have expected) for the higher levels.  

Last but not least, these methods introduced here are based in very simple concepts, which improve in efficiency the classical finite difference formalism. In addition, its numerical implementation with the  use of Mathematica is very simple. 

We have discussed the problem of the convergence of approximate solutions for the iterative Taylor method. 

The Mathieu equation as well as the equation used in the study of graphene nanotubes have both periodic coefficients.  For this reason, it has been very useful testing our methods on the Mathieu equation, before using them for the study of graphene.

\section*{Acknowledgements} 

Partial financing support is acknowledged to the Spanish MINECO (Project MTM2014-57129), the Junta de Castilla y Le\'on (Project GR224) and the Project ING 19/ i 402 of the Universidad Nacional de Rosario.

\begin{figure}
\centering
\includegraphics[width=1.0\textwidth]{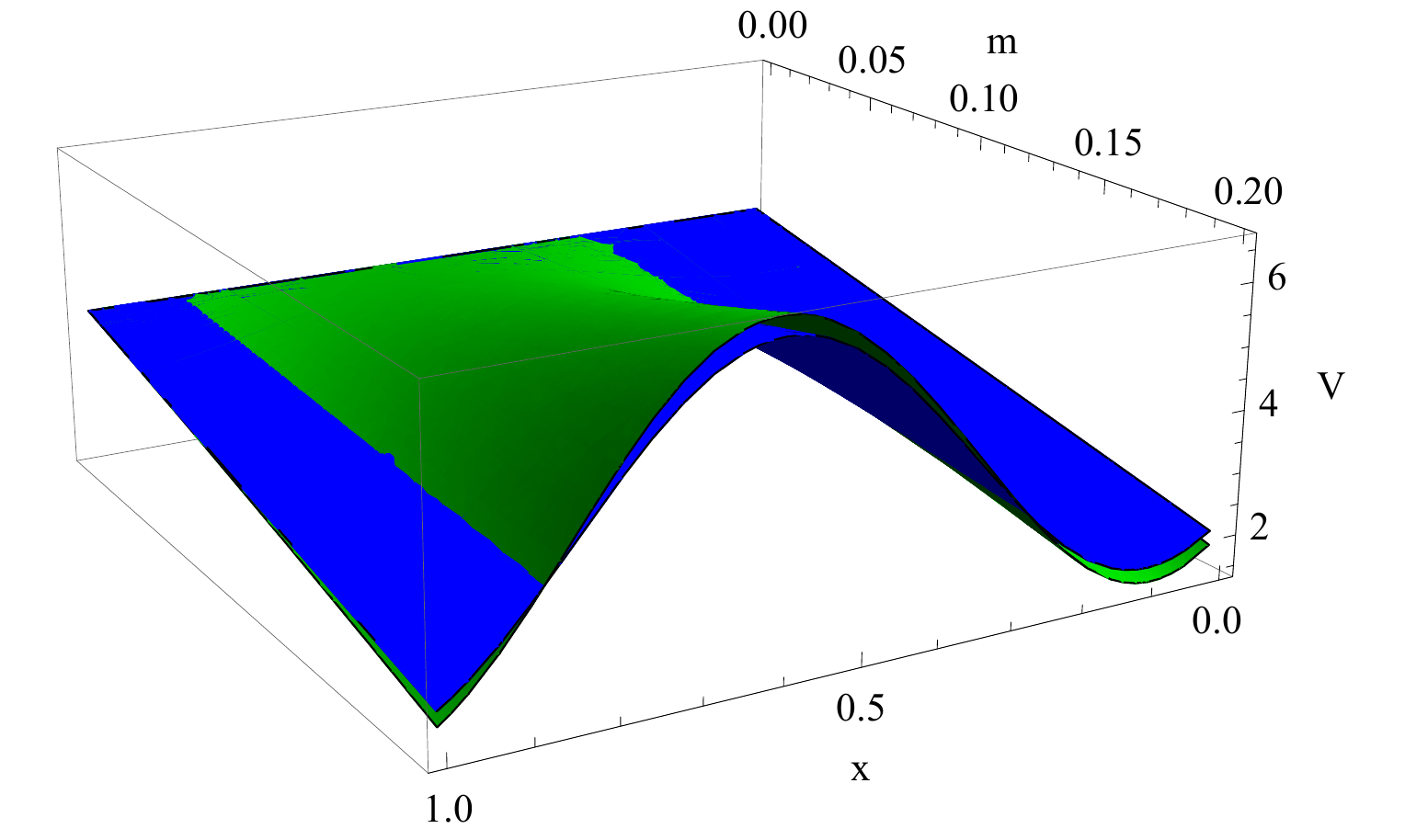}
\caption{Comparison between the exact potential (\ref{39}) and the approximate (\ref{40}), when $m<<1$ and $k_z=1$. The green and blue color correspond to the exact and approximate potentials respectively.
\label{potential}}
\end{figure}

\begin{figure}
\centering
\includegraphics[width=0.8\textwidth]{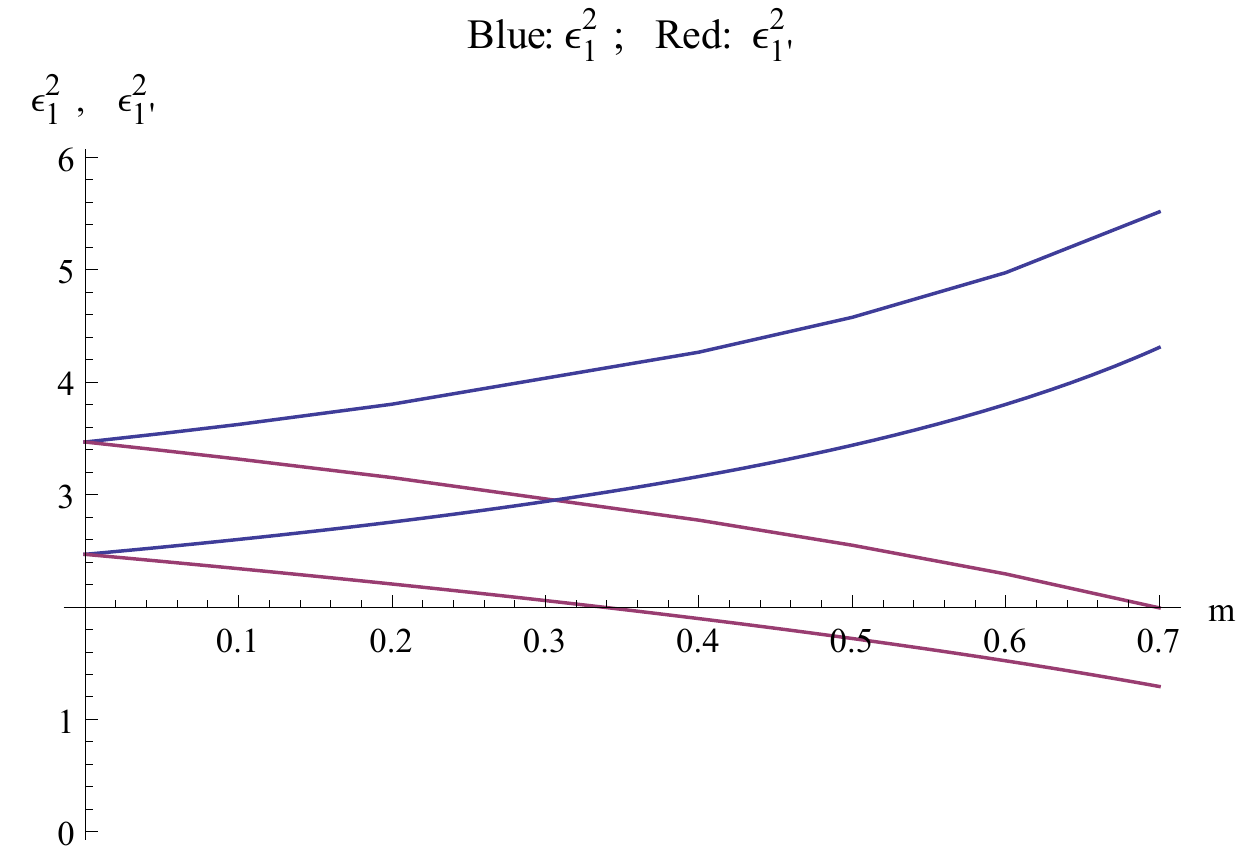}
\caption{Split between $\epsilon^2_1$ and $\epsilon^2_{1'}$ for values $m>0$. The two lower lines corresponds to $k_z=0$ and the two upper to $k_z=1$.
\label{split_1}}
\end{figure}

\begin{figure}
\centering
\includegraphics[width=0.8\textwidth]{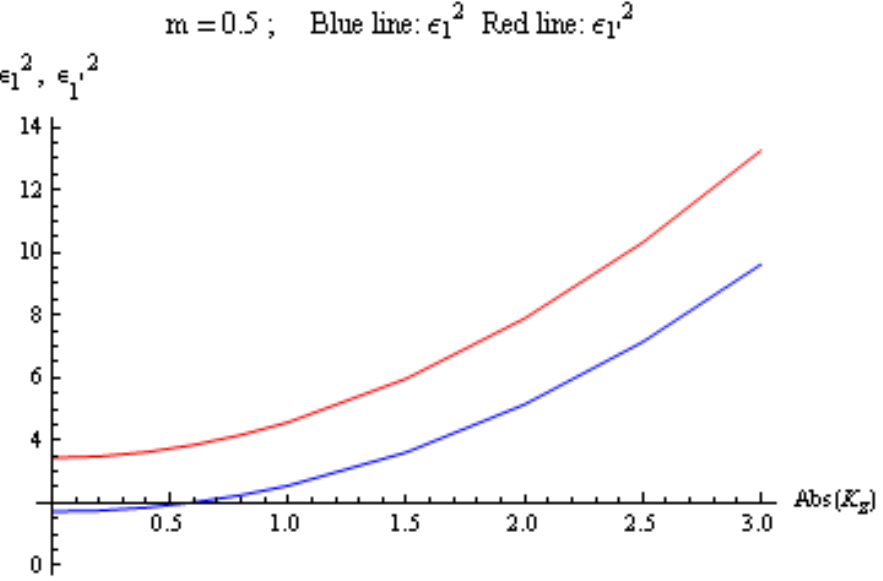}
\caption{Dependence of $\epsilon^2_1$ and $\epsilon^2_{1'}$ with $|k_z|$ for $m=0.5$ fixed.
\label{split_2}}
\end{figure}

\begin{figure}
\centering
\includegraphics[width=0.7\textwidth]{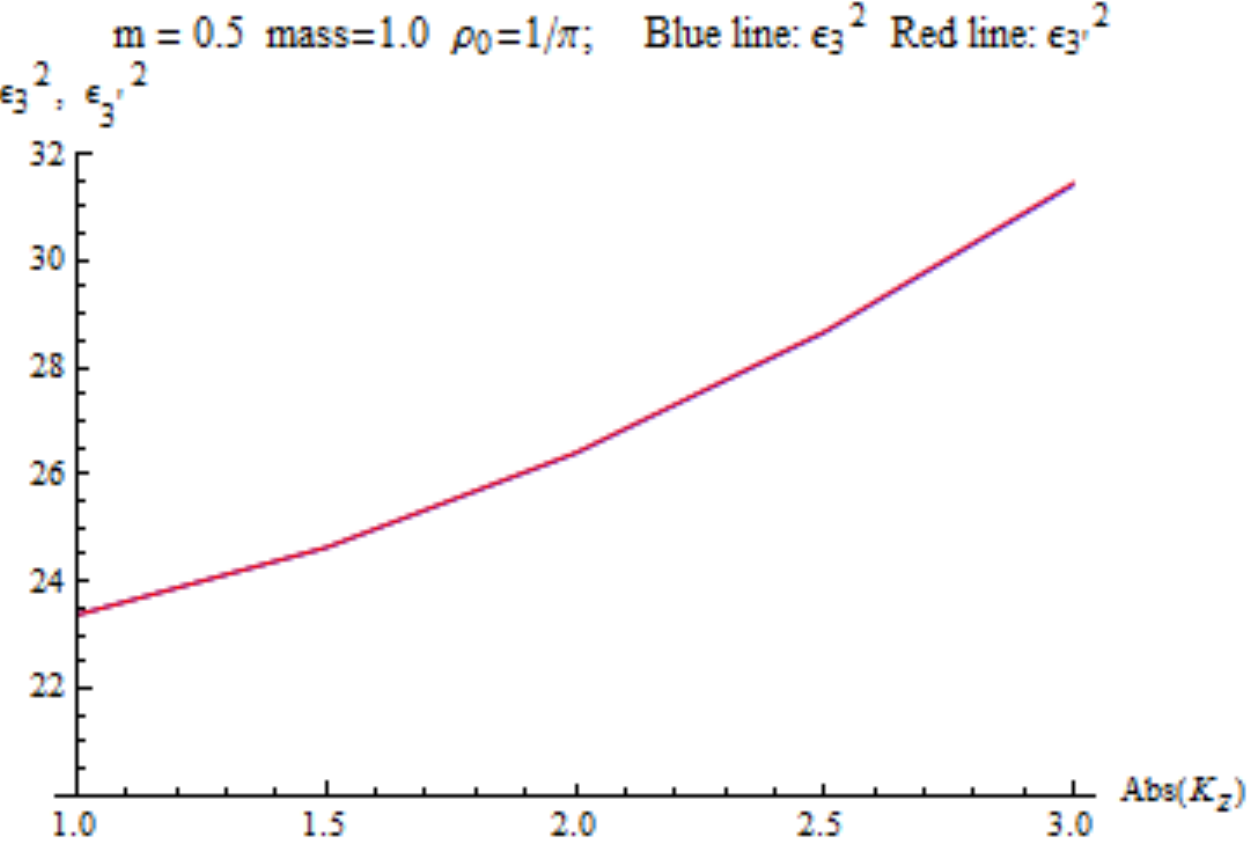}
\caption{Dependence of $\epsilon^2_3$ and $\epsilon^2_{3'}$ with $|k_z|$ for $m=0.5$ fixed.
\label{split_3}}
\end{figure}

\end{document}